\documentclass{article}
\PassOptionsToPackage{square, numbers, compress,sort}{natbib}
\usepackage[final]{nips_2017}

\usepackage[T1]{fontenc}
\usepackage{booktabs}
\usepackage{amsfonts}
\usepackage{nicefrac}
\usepackage{microtype}
\usepackage[utf8]{inputenc}
\usepackage{color}
\usepackage{graphicx}
\usepackage{algorithm,algorithmic}
\usepackage{hyperref}
\usepackage{url}
\usepackage{amsfonts}
\usepackage{amsmath}
\usepackage{booktabs}
\usepackage{wrapfig}
\usepackage[bf,small]{caption}

\renewcommand{\paragraph}[1]{\textbf{#1}}

\title{Extremely Large Minibatch SGD: \\ Training ResNet-50 on ImageNet in 15 Minutes}

\author{
  Takuya Akiba \\
  Preferred Networks, Inc. \\
  \texttt{akiba@preferred.jp}
  \And
  Shuji Suzuki \\
  Preferred Networks, Inc. \\
  \texttt{ssuzuki@preferred.jp}
  \And
  Keisuke Fukuda \\
  Preferred Networks, Inc. \\
  \texttt{kfukuda@preferred.jp}
}

\begin{document}

\maketitle
\begin{abstract}
We demonstrate that training ResNet-50 on ImageNet for 90 epochs can be achieved in 15 minutes with 1024 Tesla P100 GPUs.
This was made possible by using a large minibatch size of 32k.
To maintain accuracy with this large minibatch size, we employed several techniques such as RMSprop warm-up, batch normalization without moving averages, and a slow-start learning rate schedule.
This paper also describes the details of the hardware and software of the system used to achieve the above performance.
\end{abstract}

\section{Introduction}
Training deep neural networks is computationally expensive.
Acceleration by distributed computing is required for higher scalability
(larger datasets and more complex models) and for higher productivity (shorter training time and quicker trial and error).
This paper demonstrates that highly-parallel training is possible with a large minibatch size without losing accuracy on carefully-designed software and hardware systems.

We used the 90-epoch, ResNet-50~\cite{He2016} training on ImageNet as our benchmark.
This task has been extensively used in evaluating performance of distributed deep learning~\cite{Goyal2017, You2017, Codreanu2017}.
Table~\ref{tbl:records} shows the summary of these previous attempts along with our new results.
We achieved a total training time of 15 minutes while maintaining a comparable accuracy of 74.9\%.

The technical challenge is two-fold;
On the algorithm side,
we have to design training methods that can prevent loss of accuracy with large minibatch sizes,
while on the system side, we have to design stable and practical combinations of available hardware and software components.

{
  \tabcolsep=2.9mm
\begin{table}[h]
  \centering
  \caption{90-epoch training time and single-crop validation accuracy of ResNet-50 for ImageNet
    reported by different teams.}
  \label{tbl:records}
  \small
\begin{tabular}{c|ccc|cc} \toprule
\textbf{Team} &  \textbf{Hardware} & \textbf{Software} & \textbf{Minibatch size} & \textbf{Time} & \textbf{Accuracy} \\ \midrule
He \emph{et~al.}~\cite{He2016} & Tesla P100 $\times$ 8 & Caffe & 256 & 29 hr & 75.3 \% \\
Goyal \emph{et~al.}~\cite{Goyal2017} & Tesla P100 $\times$ 256 & Caffe2 & 8,192 & 1 hr & 76.3 \% \\
Codreanu \emph{et~al.}~\cite{Codreanu2017} & KNL 7250 $\times$ 720 & Intel Caffe & 11,520 & 62 min & 75.0 \% \\
You \emph{et~al.}~\cite{You2017} & Xeon 8160 $\times$ 1600 & Intel Caffe & 16,000 & 31 min & 75.3 \% \\
\midrule
This work & Tesla P100 $\times$ 1024 & Chainer & 32,768 & 15 min & 74.9 \% \\  \bottomrule
\end{tabular}
\end{table}

}

\section{Training Procedure for Large Minibatches}
\label{sec:training}

We build on the training procedure proposed by \cite{Goyal2017},
and the same settings are used unless otherwise specified.
We briefly highlight the differences in this section.
For further details, please see Appendix~\ref{sec:details}.

\paragraph{RMSprop Warm-up.}
We found that the primary challenge is the optimization difficulty at the start of training.
To address this issue,
we start the training with RMSprop~\cite{Tieleman2012}, then gradually transition to
SGD.

\paragraph{Slow-Start Learning Rate Schedule.}
To further overcome the initial optimization difficulty,
we use a slightly modified learning rate schedule
with a longer initial phase and lower initial learning rate.

\paragraph{Batch Normalization without Moving Averages.}
With the larger minibatch sizes, the batch normalization moving averages of the mean and variance became inaccurate estimates of the actual mean and variance.
To cope with this problem, we only considered the last minibatch, instead of the moving average,
and used all-reduce communication on these statistics to obtain the average over all workers before validation.

\section{Software and Hardware Systems}

\paragraph{Software.}
We used \emph{Chainer}~\cite{Tokui2015} and
\emph{ChainerMN}~\cite{Akiba2017}.  Chainer is an open-source
deep learning framework featuring the define-by-run
approach.  ChainerMN is an add-on package for Chainer enabling multi-node
distributed deep learning with synchronous data-parallelism.
We used development branches based on versions 3.0.0rc1 and 1.0.0, respectively.
As the underlying communication libraries, we used NCCL
version 2.0.5 and Open MPI version
1.10.2.
While computation was generally done in single precision,
in order to reduce the communication overhead during all-reduce operations,
we used half-precision floats for communication.
In our preliminary experiments, we observed that the effect from using half-precision
in communication on the final model accuracy was relatively small.

\paragraph{Hardware.}
We used \emph{MN-1}, an in-house cluster owned by Preferred Networks, Inc.~designed to facilitate research and development of deep learning.
It consists of 128 nodes, where each node has two Intel Xeon E5-2667 processors (3.20 GHz, eight cores), 256 GB memory and eight NVIDIA Tesla P100 GPUs.
The nodes are interconnected by Mellanox Infiniband FDR.

\section{Experimental Results}

\begin{wrapfigure}{r}{5.0cm}
  \centering
  \vspace{-1em}
  \includegraphics[width=1 \hsize]{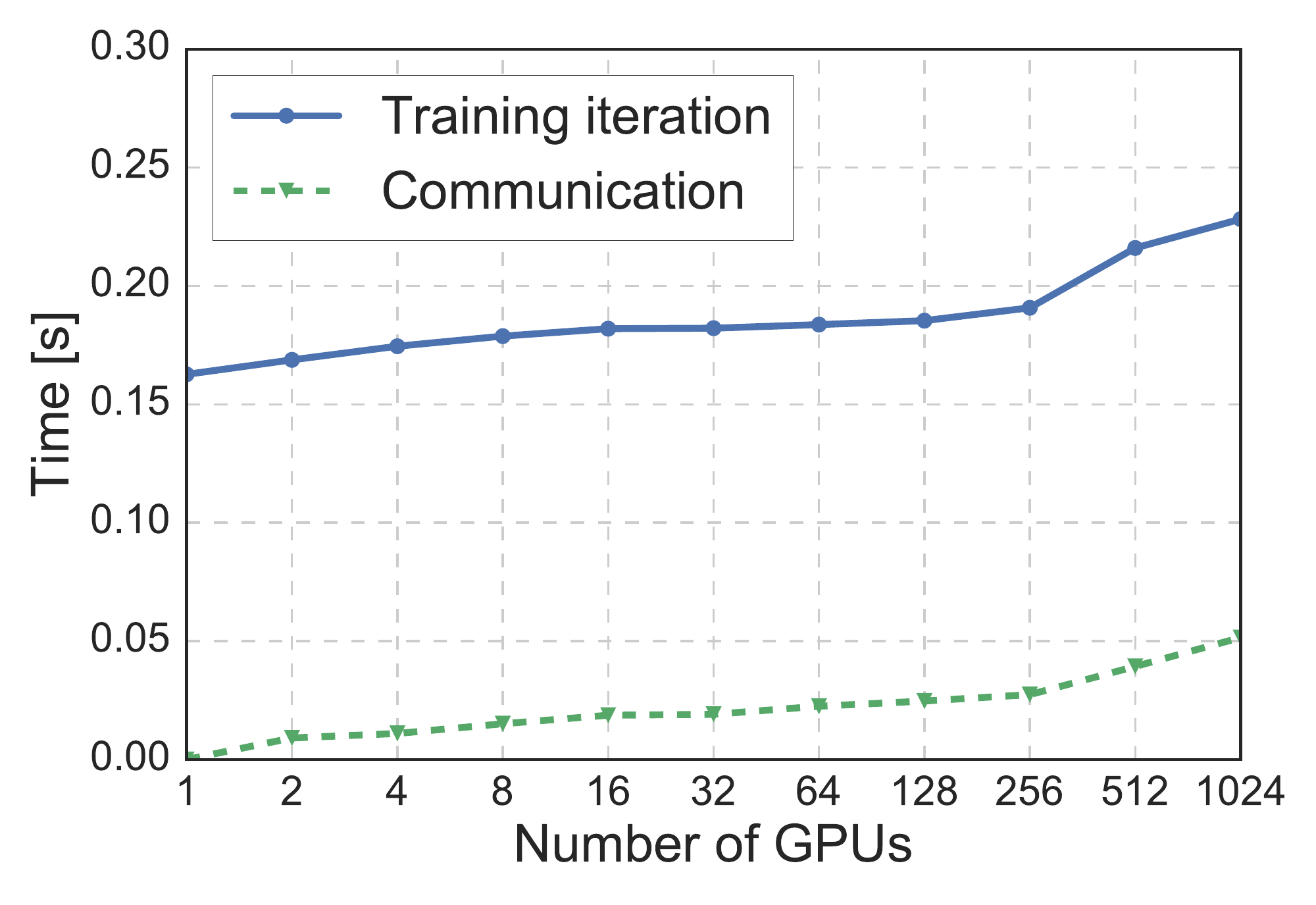}
  \caption{Iteration and communication time for different numbers of GPUs.}
  \label{fig:scaling}
  \vspace{-1em}
\end{wrapfigure}

For running time and accuracy, the mean and standard deviation from five independent runs are reported.
The per-worker minibatch size was 32, and the total minibatch size was 32k with 1024 workers.

\paragraph{Training Time.}
Using 1024 GPUs,
the training time was $897.9 \pm 3.3$ seconds for 90 epochs, including validation after each epoch.
Figure~\ref{fig:scaling} illustrates
the average communication time (i.e., all-reduce operations)
and time to complete a whole iteration (i.e., forward and backward computation, communication, and optimization)
over 100 iterations.
Our scaling efficiency when using 1024 GPUs
is 70\% and 80\% in comparison to single-GPU and single-node (i.e., 8 GPUs) baselines, respectively.

\paragraph{Accuracy.}
After training on 90 epochs using 1024 GPUs with the training procedure designed in Section~\ref{sec:training},
the top-1 single-crop accuracy on the validation images was $74.94\% \pm 0.09$.
As we can observe from Table~\ref{tbl:records},
this accuracy is comparable to that of previous results using ResNet-50.
Therefore, it shows that
ResNet-50 can be trained on ImageNet with a minibatch size of 32k
without severely degrading the accuracy,
which validates our claim that training of ResNet-50 can be successfully completed in 15 minutes.

\section*{Acknowledgements}
{
\small
The authors thank
Y.~Doi, G.~Watanabe, R.~Okuta, T.~Kikuchi, and M.~Sakata for help on experiments,
T.~Miyato and S.~Tokui for fruitful discussions, and
H.~Maruyama, R.~Calland, and C.~Loomis for helping to improve the manuscript.
}

{
\small
\bibliographystyle{abbrv}
\bibliography{main}
}

\appendix

\section{Details of Training Procedure}
\label{sec:details}

\subsection{RMSprop Warm-up}
Our update rule is a simple combination of momentum SGD and RMSprop~\cite{Tieleman2012} (a variant with momentum), defined as follows:
\begin{align*}
m_t &= \mu_2 m_{t-1} + (1 - \mu_2) g_t^2, \\
\Delta_{t} &= \mu_1 \Delta_{t-1} - \left( \alpha_\text{SGD} + \frac{\alpha_\text{RMSprop}}{\sqrt{m_t} + \varepsilon}  \right) g_t, \text{and} \\
\theta_{t} &= \theta_{t-1} + \eta \Delta_{t}.
\end{align*}
Here, $t$ denotes the current index of iteration.
The weights, gradients, momentum, and moving average of the second moment of the gradient at the $i$-th iteration are represented by $\theta_i, g_i, \Delta_i$, and $m_i$ respectively.
The inputs are $g_t, \theta_{t-1}, \Delta_{t-1}$, and $m_{t-1}$,
and the outputs are $\theta_{t}, \Delta_{t}$, and $m_t$.
Hyperparameters are $\eta$, $\mu_1$, $\mu_2$, $\varepsilon$, $\alpha_\text{SGD}$ and $\alpha_\text{RMSprop}$:
$\eta$ is the learning rate,
$\mu_1$ determines the amount of momentum,
$\mu_2$ is the coefficient for the moving average of the gradient second moment,
and $\varepsilon$ is a small number added for numerical stability.
We used $\mu_1 = 0.9, \mu_2=0.99$, and $\varepsilon=10^{-8}$ throughout our experiments.
Parameters $\alpha_\text{SGD}$ and $\alpha_\text{RMSprop}$
determine the balance between momentum SGD and RMSprop:
when $\alpha_\text{RMSprop}=0$,
it corresponds to the standard momentum SGD,
and when $\alpha_\text{SGD}=0$, it matches RMSprop.

We start with RMSprop (i.e., $\alpha_\text{SGD} \approx 0$), and then smoothly switch to SGD (i.e., $\alpha_\text{SGD} = 1$).
For the transition schedule,
we use a function that is similar to the exponential linear unit (ELU) activation function~\cite{Clevert2015} defined as follows:
\begin{align*}
  \alpha_\text{SGD} = \begin{cases}
    \frac12 {\exp(2(\text{epoch} - \beta_\text{center}) / \beta_\text{period})} & (\text{epoch} < \beta_\text{center})\\
    \frac12 + 2(\text{epoch} - \beta_\text{center}) / \beta_\text{period} & (\text{epoch} < \beta_\text{center} + \frac12 \beta_\text{period})\\
    1 & (\text{otherwise})\\
  \end{cases}.
\end{align*}
Here, $\beta_\text{center}$ and $\beta_\text{period}$ are hyperparameters.
First, $\alpha_\text{SGD}$ increases exponentially.
At the $\beta_\text{center}$-th epoch, $\alpha_\text{SGD}$ reaches $\frac12$.
After that, it increases linearly until the $\beta_\text{center} + \frac12 \beta_\text{period}$-th epoch.
At the $\beta_\text{center} + \frac12 \beta_\text{period}$-th epoch, $\alpha_\text{SGD}$ becomes 1,
and we set $\alpha_\text{SGD}=1$ for the remainder of the training.
We set $\beta_\text{center}=10$ and $\beta_\text{period}=5$ throughout our experiments.

We used $\eta_\text{RMSprop} = 0.0003$ for the learning rate of RMSprop.
Let $\eta_\text{SGD}$ be the learning rate of SGD, which will be discussed in the next subsection.
To incorporate different learning rates of SGD and RMSprop,
we set $\eta = \eta_\text{SGD}$
and $\alpha_\text{RMSprop} = (1 - \alpha_\text{SGD}) \eta_\text{RMSprop} / \eta_\text{SGD}$.
One might think that the rule would be simpler
if we multiply $\eta_\text{SGD}$ to $\alpha_\text{SGD}$ beforehand,
but we should make $\Delta_{t}$ independent from varying learning rates
for momentum correction proposed by Goyal \emph{et al.}~\cite{Goyal2017}.

A method similar to our RMSprop warm-up is used by Wu \emph{et al.}~\cite{Wu2016} for a machine translation task.
They use the Adam~\cite{Kingma2014} optimizer at the beginning, then switch to SGD.
In our preliminary experiments, we found that RMSprop performs better for our task.
In addition, Wu \emph{et al.} suddenly switches from Adam to SGD.
However, we found that sudden transition severely impacts training and has a negative effect on the final results.
Therefore, we designed a smooth transition from RMSprop to SGD.
We examined a few transition functions including linear and sigmoid functions.
Linear functions have a similar problem at the beginning of the transition.
ELU and sigmoid performed similarly, but ELU performs slightly better, so we opted for ELU.

\subsection{Slow-Start Learning Rate Schedule}

\newcommand{\lr}{\eta_\text{base}}

Let $\lr$ be the initial learning rate under the linear rule by Goyal \textit{et al.}~\cite{Goyal2017}.
Specifically, $\lr= 0.1 \cdot \frac{b_\text{total}}{256} = 0.1 \cdot \frac{n b_\text{local}}{256}$,
where $n$ is the number of workers,
$b_\text{local}$ is the local batch size for each worker,
and $b_\text{total}$ is the total batch size among all workers (i.e., $b_\text{total} = n b_\text{local}$).
In our experiments, $n=1024$ and $b_\text{local}=32$, and thus $\lr=12.8$.
Goyal \emph{et~ al.}'s learning rate schedule is as follows: $\lr$ for first 30 epochs, $0.1 \cdot \lr$ for the next 30 epochs, $0.01 \cdot \lr$ for the following 20 epochs, and $0.001 \cdot \lr$ for the last 10 epochs.

To overcome the initial optimization difficulty,
we used a slow-start schedule;
our learning rate for SGD was
$0.5 \cdot \lr$ for the first 40 epochs,
$0.075 \cdot \lr$ for the next 30 epochs,
$0.01 \cdot \lr$ for the following 15 epochs,
and $0.001 \cdot \lr$ for the last 5 epochs.

\end{document}